\documentclass[aps,prl,showpacs,superscriptaddress,groupedaddress]{revtex4}  
\usepackage{graphicx}  
\usepackage{dcolumn}   
\usepackage{bm}        
\usepackage{amssymb}   
\usepackage{subfigure}
\usepackage{sidecap}

\begin{document}

\title{Detection-Loophole-Free Test of Quantum Nonlocality, and Applications}
\author{B. G. Christensen} \affiliation{Department of Physics, University of Illinois at Urbana-Champaign, Urbana, IL 61801, USA.}
\author{K. T. McCusker} \affiliation{Department of Physics, University of Illinois at Urbana-Champaign, Urbana, IL 61801, USA.}
\author{J. B. Altepeter} \affiliation{Department of Physics, University of Illinois at Urbana-Champaign, Urbana, IL 61801, USA.}
\author{B. Calkins} \affiliation{National Institute of Standards and Technology, Boulder, CO 80305, USA.}
\author{T. Gerrits} \affiliation{National Institute of Standards and Technology, Boulder, CO 80305, USA.}
\author{A. E. Lita} \affiliation{National Institute of Standards and Technology, Boulder, CO 80305, USA.}
\author{A. Miller} \affiliation{National Institute of Standards and Technology, Boulder, CO 80305, USA.} \affiliation{Department of Physics, Albion College, Albion, MI 49224, USA.}
\author{L. K. Shalm} \affiliation{National Institute of Standards and Technology, Boulder, CO 80305, USA.}
\author{Y. Zhang} \affiliation{National Institute of Standards and Technology, Boulder, CO 80305, USA.} \affiliation{Department of Physics, University of Colorado at Boulder, CO 80309, USA.}
\author{S. W. Nam} \affiliation{National Institute of Standards and Technology, Boulder, CO 80305, USA.}
\author{N. Brunner} \affiliation{D\'epartement de Physique Th\'eorique, Universit\'e de Gen\'eve, 1211 Gen\'eve, Switzerland.} \affiliation{H. H. Wills Physics Laboratory, University of Bristol, Tyndall Avenue, Bristol, BS8 1TL, United Kingdom.}
\author{C. C. W. Lim} \affiliation{Group of Applied Physics, Universit\'e de Gen\'eve, 1211 Gen\'eve, Switzerland.}
\author{N. Gisin} \affiliation{Group of Applied Physics, Universit\'e de Gen\'eve, 1211 Gen\'eve, Switzerland.}
\author{P. G. Kwiat} \affiliation{Department of Physics, University of Illinois at Urbana-Champaign, Urbana, IL 61801, USA.}

\date{\today}

\begin{abstract}
We present a source of entangled photons that violates a Bell inequality free of the ``fair-sampling" assumption, by over 7 standard deviations.  This violation is the first reported experiment with photons to close the detection loophole, and we demonstrate enough ``efficiency" overhead to eventually perform a fully loophole-free test of local realism.  The entanglement quality is verified by maximally violating additional Bell tests, testing the upper limit of quantum correlations.  Finally, we use the source to generate ``device-independent" private quantum random numbers at rates over 4 orders of magnitude beyond previous experiments.
\\
\\
\noindent{This document has been published at http://prl.aps.org/abstract/PRL/v111/i13/e130406 in \textit{Phys. Rev. Lett.}}
\end{abstract}

\maketitle

In 1935, Einstein, Podolsky, and Rosen suggested that certain quantum mechanical states must violate one or both of the fundamental classical assumptions of \textit{locality} (sufficiently distant events cannot change the outcome of a nearby measurement) and \textit{realism} (the outcome probabilities of potential measurements depend only on the state of the system). These \textit{nonclassical} two-particle states exhibit multiple-basis correlations (or anti-correlations), and are referred to as ``entangled".  Because locality and realism are so fundamental to classical intuition, a central debate in 20th century physics \cite{1} revolved around the following question: could an alternative to quantum mechanics---a local realistic theory---explain entanglement’s seemingly nonclassical correlations? In 1964, John Bell devised a way to in principle answer this question experimentally, by analyzing the limit of allowed correlations between measurements made on an ensemble of any classical system \cite{2}.  If performed under sufficiently ideal conditions, a violation of Bell’s inequality would \textit{conclusively rule out all possible local realistic theories}.  Although entanglement has been experimentally demonstrated and the Bell inequality violated in a myriad of non-ideal experiments \cite{3,4,5,6,7,8,9,10,11,12}, each of these experiments fails to overcome at least one of two critical obstacles.

The first obstacle–--the ``locality loophole"–--addresses the possibility that a local realistic theory might rely on some type of signal sent from one entangled particle to its partner (e.g., a signal containing information about the specific measurement carried out on the first particle), or from the measurement apparatus to the source (known as the freedom of choice loophole).  These loopholes have thus far only been closed using entangled photons \cite{8,freedomofchoice}; photons traveling in different directions can be measured at places and times which are \textit{relativistically} strictly simultaneous (i.e., in a space-like separated configuration).  The second obstacle---the ``detection loophole"---addresses the fact that even maximally entangled particles, when measured with low-quantum-efficiency detectors, will produce experimental results that can be explained by a local realistic theory.  To avoid this, almost all previous experiments have had to make ``fair sampling" assumptions that the collected photons are ``typical" of those emitted (this assumption is demonstrably \textit{false} \cite{13} for many of the pioneering experiments using atomic cascades \cite{3,4}, and has been intentionally exploited to fake Bell violations in recent experiments \cite{14}).  The detection loophole has been closed in several matter systems: ions \cite{10}, superconductors \cite{11}, and atoms \cite{12}, whereas high-efficiency tests with photons have been lacking until very recently \cite{15}.  Unfortunately, the results presented in \cite{15} are actually susceptible to multiple loopholes (in addition to the locality loophole) \cite{16,17}.  Specifically, their use of a source without well-defined individual experimental trials has a loophole where the ambiguous definition of a coincidence count can be exploited to produce a Bell violation, even for a completely local realistic hidden-variable model (see Supplemental Information \cite{18} for more information).  In addition, the lack of a random measurement basis selection requires a fair-sampling assumption, since temporal drifts in pair production or detection rates, or detector latency, can also lead to a Bell violation without entanglement \cite{18}.  Although these issues may seem pedantic, the point of a loophole-free Bell test is to definitively rule out any local realistic theory, which becomes all the more critical as Bell tests are used to ``certify" quantum communications.

Here we report the first experiment that fully closes the detection loophole with photons, which are then the only system in which both loopholes have been closed, albeit not simultaneously.  Moreover, we show that the source quality is high enough to provide the best test to date of the quantum mechanics prediction itself.  Finally, we apply the stronger-than-classical correlations to verify the creation of true random numbers, achieving rates over 4 orders of magnitude beyond all past experiments.

The first form of a Bell inequality that was experimentally feasible and did not require assumptions such as ``fair sampling" was the Clauser-Horne (CH) inequality \cite{19}, which places the following constraints for any local realistic theory:
\begin{eqnarray}
p_{12}(a,b)+p_{12}(a,b')+p_{12}(a',b)-p_{12}(a',b') \leq p_{1}(a)+p_{2}(b)
\label{eq:one}
\end{eqnarray}
where $a, a' (b, b')$ are the settings for detector 1 (2), $p_{1(2)}(x)$ denotes the probability of a count for any given trial at detector 1 (2) with setting $x$, and $p_{12}(x,y)$ denotes the probability of a coincidence count with settings x and y for detectors 1 and 2, respectively.  The inequality can be violated using maximally entangled states (e.g.,$\left (|HH \rangle + |VV \rangle \right )/\sqrt{2}$, where H and V represent the polarization of the photons), assuming a detection efficiency $\eta > 2(\sqrt{2}-1)\approx 0.828$; this is the lower efficiency limit for any maximally entangled two-particle system measured with a pair of detectors that each has two settings \cite{upperlimitusingmaxentangle}.  However, further analysis by Eberhard \cite{20} showed that with non-maximally entangled states, e.g., $|\psi _{r} \rangle=\left (r|HH \rangle + |VV \rangle \right )/\sqrt{1+r^{2}}$, the detector efficiency requirement could be reduced to $2/3$, although the tolerable amount of background counts in the detector is very small in this limit.  Essentially, using a small value of $r$, one can choose $a$ and $b$ to nearly block the vertically polarized single counts (thereby decreasing the RHS of Eq. 1), while choosing $a'$ and $b'$ to maximize the LHS \cite{18}.  For the background levels in our experiment, a value of $r = 0.26$ allows us to maximally violate the CH inequality.

In order to determine the probabilities in Eq. 1, we normalize the measured singles and coincidence rates to the number of trials with the specific analyzer setting for each term. We can then write
\begin{eqnarray}
B=\frac{C_{12}(a,b)}{N(a,b)}+\frac{C_{12}(a,b')}{N(a,b')}+\frac{C_{12}(a',b)}{N(a',b)}-\frac{C_{12}(a',b')}{N(a',b')}-\frac{S_{1}(a)}{N(a)}-\frac{S_{2}(b)}{N(b)} \leq 0
\label{eq:two}.
\end{eqnarray}
where $C_{12}(x,y)$ are the coincidence counts, and $S_{1}(x)$ the singles counts for the duration of the experiment, $N(x,y)$ is the total number of trials where the detectors settings were $x,y$, and $N(x)$  is the number of trials where the channel setting was $x$ (regardless of the setting on the other side).

In order to avoid the coincidence-time loophole \cite{16,17,18} (one of the same loopholes present in the reported data for the previous photon experiment \cite{15}), we use a Pockels cell between crossed polarizers to periodically transmit short ``bursts" of the pump laser.  Each burst corresponds to a single well-defined ``event", easily distinguished with the detectors.  Care must still be taken, however, to guarantee that there is no temporally-correlated effect that unduly affects the measured counts.  For example, laser power drift can lead to a violation with non-entangled photons, if the order of the measurements is not made randomly \cite{18}.  We address this issue by measuring each of the terms in Eq. 2 multiple times while randomly choosing the detector settings, and then determine the counts and relative errors (due to both finite counting statistics as well as multiple measurements of each term).

For our entanglement source \cite{18}, we focus the 355-nm pulsed laser onto two orthogonal nonlinear crystals to produce polarization-entangled photon pairs at 710 nm, via spontaneous parametric downconversion \cite{21}.  The degree of entanglement of the downconverted photons can be controlled using waveplates to manipulate the pump polarization, i.e., a pump polarization of $\left (|V \rangle + r e^{i \phi}|H \rangle \right )/\sqrt{2}$ will produce the entangled state $\left (|HH \rangle + r e^{i (\phi+\phi _{c})}|VV \rangle \right )/\sqrt{2}$, where $\phi _{c}$ is the relative phase picked up in the nonlinear crystals \cite{22}.

\begin{figure}
\includegraphics[scale=0.33]{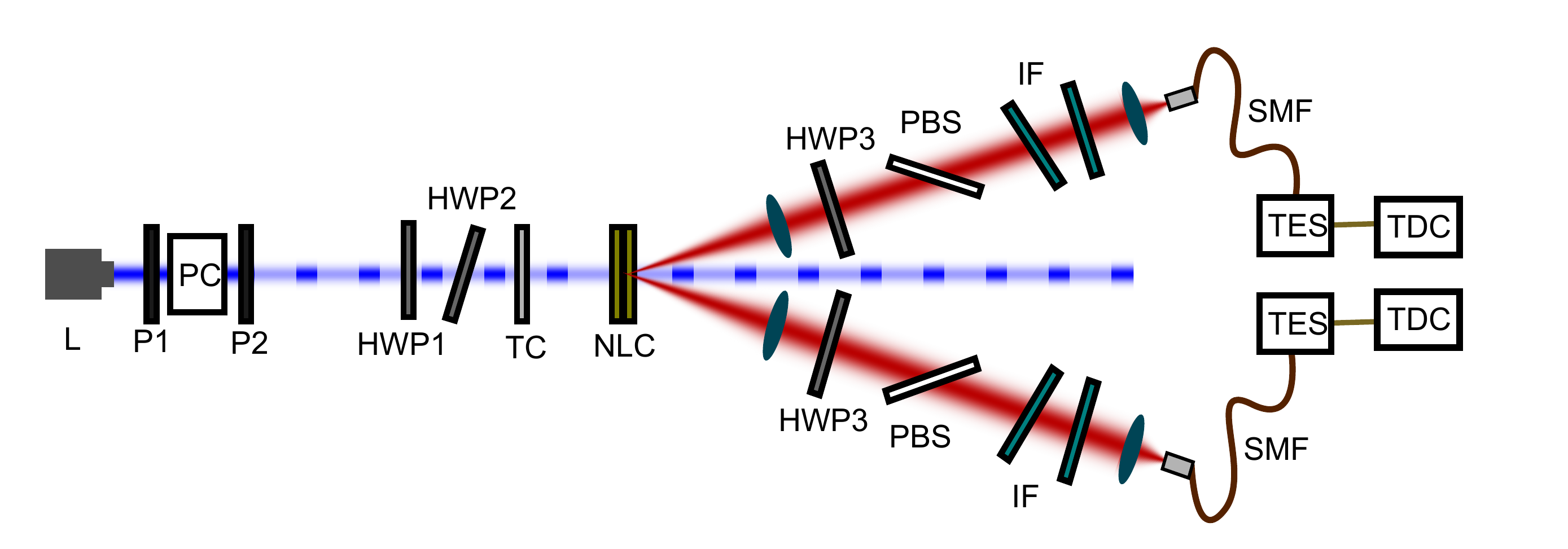}
\caption{\label{fig:first} A diagram of the system used to violate the CH Bell inequality \cite{18}.  We ``pulse" our laser (L) by putting a Pockels cell (PC) between crossed polarizers (P1 and P2).  The Pockels cell is periodically turned on for a short time to allow 240 laser pulses to transmit through P2, thus creating ``event intervals" that the detectors can distinguish, which we can then use as well-defined trials.  Downconversion is produced in paired nonlinear BiBO crystals (NLC).  The produced state is controlled through the half-wave plates HWP1 and HWP2, which control the relative amplitude and phase of the $|HH \rangle$ and $|VV \rangle$ downconversion terms. We attain very high entanglement quality by compensating temporal decoherence caused by group-velocity dispersion in the downconversion crystals \cite{36}, using a BBO crystal (TC).  The downconversion creation spot is imaged onto a single-mode fiber (SMF).  HWP3 sets the basis for the polarization analysis (based on input from QRNG data) by the Brewster’s angle polarizing beam splitter (PBS).  Custom spectral interference filters (IF) are used to only detect spectrally conjugate photons.  Finally, the photons are detected by transition-edge-sensor (TES) detectors; the output signals are sent to a time-to-digital converter (TDC) to record all detection event times.}
\end{figure}

The state that maximally violates a CH Bell inequality is a compromise between non-unit system efficiency (which pushes to smaller $r$ values) and non-zero (unpolarized) background counts (contributing to the singles rate, which limits the minimum usable $r$ value) \cite{20}.  We model our system to find the ideal state and analysis settings, based on our measured background counts and efficiency \cite{18}.  The polarization correlations for the Bell test are measured with a polarization analyzer consisting of a fixed Brewster’s angle polarizing beam splitter (we detect the transmitted photons), preceded by a motion-controlled anti-reflection (AR) coated half-wave plate to choose the basis of the projective measurement.  In the experiment we randomly choose the measurement settings ($a$ or $a'$, $b$ or $b'$) using the output from a separate photon-arrival-time-based quantum random number generator (QRNG) \cite{23}.

The energy correlations in the daughter photons ($\omega _{p}=\omega _{i}+\omega _{s}$) allow us to spectrally filter the photons to ensure the collection of the conjugate photons.  To do so, we use custom tunable 20-nm interference filters \cite{18}, centered on twice the pump wavelength (710 nm), achieving an estimated $95\%$ spectral heralding efficiency (the detection efficiency if there were \textit{no} other losses in the system).  The Gaussian spatial-mode of the pump allows the momentum correlations of the daughter photons to be filtered with a single-mode fiber (SMF) \cite{24}.  Here, the collection of one photon into a SMF heralds a photon, to a very high approximation, in a well-defined Gaussian mode on the opposite side of the downconversion cone.  We are able to collect the conjugate photon in its own SMF with an estimated $90\%$ efficiency (assuming no other losses).  The SMF is then fusion-spliced to a fiber connected to a TES detector \cite{25}.

These detectors are made using a thin tungsten film embedded in an optical stack of materials to enhance the absorption \cite{26}.  Photons are delivered to the detector stack using a SMF for 1550 nm, which is AR-coated for 710 nm and fusion-spliced (with less than $5\%$ loss) to the 710-nm SMF used for downconversion collection.  The detectors, cooled to \url{~}100 mK using an adiabatic demagnetization refrigerator, are voltage-biased at their superconducting transition so that absorbed photons cause a measurable change in the current flowing through the tungsten film.  The change in current is measured with a superconducting quantum interference device amplifier, the output of which is connected to room-temperature electronics for processing before being sent to a time-to-digital converter (TDC).  The stream of timetags from the TDC are sent to a computer and saved for later analysis.
Accounting for spectral and spatial filtering, detector efficiency, \url{~}7\% from all other transmission losses, we arrive at a final detection efficiency of $75\% \pm 2\%$, sufficient to violate a Bell inequality without needed any extra fair-sampling assumption. We collected timetags for the Bell test in ``blocks" of 1 second (25,000 trials per second) at each measurement setting, for a total of 4450 blocks.  The data, summarized in Table 1, shows a $7.7$-$\sigma$ violation of the CH inequality (Eq. 2): $B = 5.4$x$10^{-5} \pm 7.0$x$10^{-6}$ \cite{18}  Our results are in good agreement with those predicted using our measured entangled state, after accounting for the measured background and fluorescence noise.

\begin{table}
\caption{\label{tab:table1}The accumulated measurements.  We used the settings $a = 3.8^{o}$, $a' = -25.2^{o}$, $b = -3.8^{o}$, $b' = 25.2^{o}$, and $r = 0.26$.  We cycled through these measurement settings randomly (using QRNG data \cite{23} to pick the basis for a given run), changing the measurement settings in intervals of 1 second, with 25,000 trials in each 1 second interval.}
\begin{ruledtabular}
\begin{tabular}{lcccr}
Settings & Singles(A) & Coincidences & Singles(B) & Trials\\
\hline
$a,b$ & 46,068 & 29,173 & 46,039 & 27,153,020\\
$a,b'$ & 48,076 & 34,145 & 146,205 & 28,352,350\\
$a',b$ & 150,840 & 34,473 & 47,447 & 27,827,318\\
$a',b'$ & 150,505 & 1,862 & 144,070 & 27,926,994\\
\end{tabular}
\end{ruledtabular}
\end{table}

It is also informative to normalize the Bell inequality in a slightly different manner:
\begin{eqnarray}
B'=\frac{p_{12}(a,b)+p_{12}(a,b')+p_{12}(a',b)-p_{12}(a',b')}{p_{1}(a)+p_{2}(b)} \leq 1
\label{eq:three}.
\end{eqnarray}
In this form, the Bell parameter $B'$ is proportional to the system efficiency.  We achieve $B' = 1.015\pm 0.002$, implying a $1.5\%$ efficiency overhead (i.e., we could tolerate $1.5\%$ more loss in each channel, with the current level of background).  This is important for both a fully loophole-free test of Bell's inequality (requiring larger separation of the detectors to close the locality loophole), as well as for device-independent quantum information protocols.  In both cases, active switch elements are necessary, as the introduction of a beam splitter to make an analysis basis choice would re-open the detection loophole \cite{27}.  With the current efficiency overhead, we can support such devices, e.g., Pockels cells, which tend to be slightly lossy optical elements ($\sim 1\%$).

\begin{figure}
\includegraphics[scale=0.25]{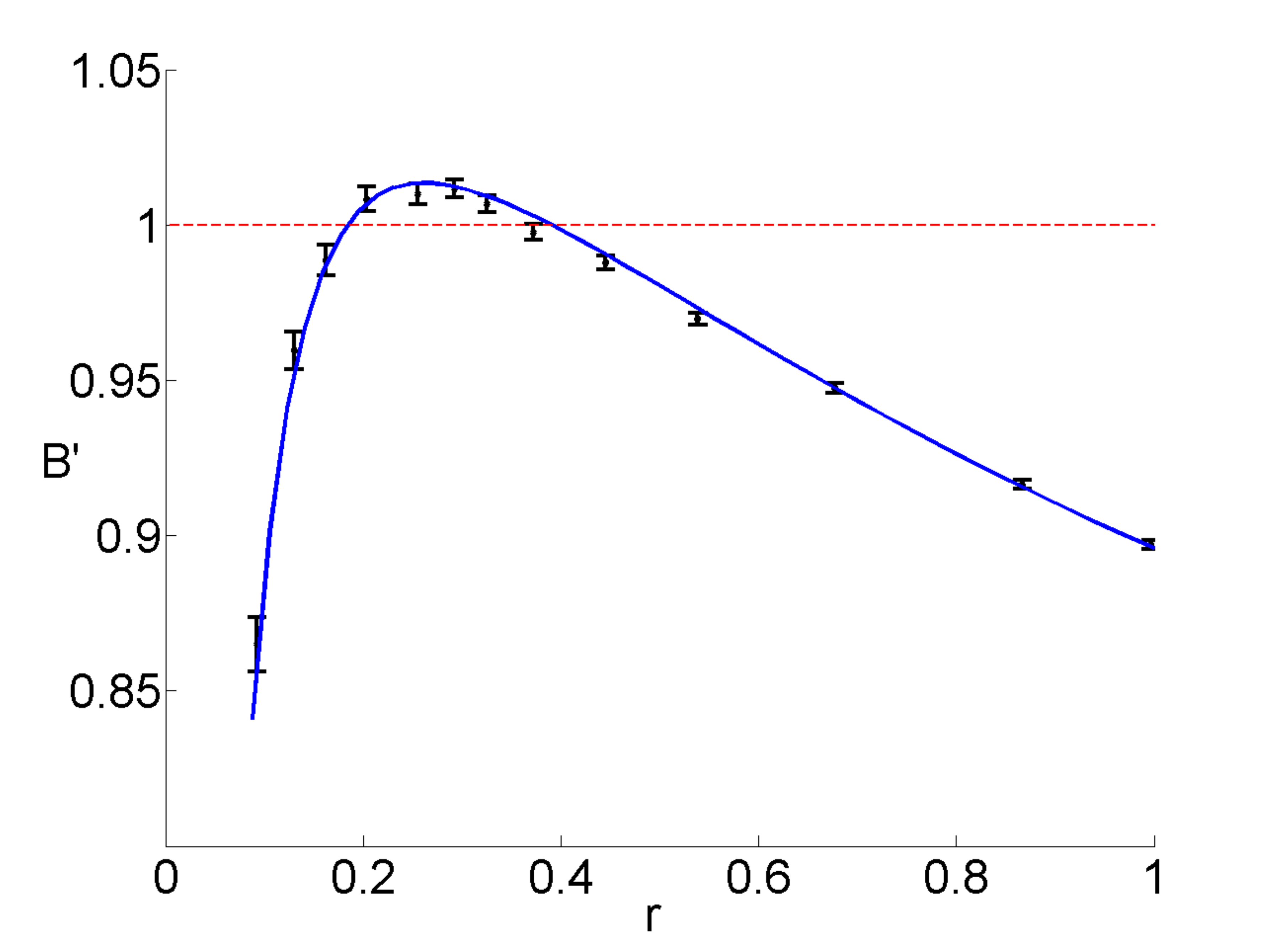}
\caption{\label{fig:second} A plot of the Bell parameter $B'$ (Eq. 3) as a function of the produced entangled state.  $B' > 1$ (red dashed line) is not possible for any local realistic theory.  Data points in black are the measured $B'$ as $r$ is varied in the state $\left (r|HH \rangle + |VV \rangle \right )/\sqrt{1+r^{2}}$; $r = 1$ ($0$) corresponds to a maximally entangled (separable) state.  For this plot, every data point was measured for 30 seconds at each measurement setting; the particular settings were optimally chosen based on the model of our source for each value of $\theta$.  The blue line represents the Bell parameter we expect from the model of our source.  Here, to improve the statistics, we did not pulse our source with the Pockels cell.  We see violations for $0.20 < r < 0.33$.}
\end{figure}

The state $|\psi _{r=0.26} \rangle$ that achieves the most statistically significant violation of local realism for our system is actually weakly entangled, with a measured concurrence of only 0.49.  This low concurrence is an expected characteristic of the type of high-purity nonmaximally entangled state best suited for violating the CH inequality with non-ideal detectors, but belies the unprecedented quality of the experimental apparatus used to generate the state itself.  We thus reconfigured the experimental apparatus shown in Fig. 1 to produce a variety of high-purity, high-fidelity quantum states between totally separable and maximally entangled; see Fig. 2 for a plot of the CH inequality violation as a function of state separability.  When configured for maximal entanglement, the source produces a state with $99.7 \pm 0.05\%$ ($99.5 \pm 0.05\%$) visibility in the H/V (H+V/H-V) basis, and a canonical CHSH Bell violation \cite{28} of $2.827 \pm 0.017$--within error of the maximum violation allowed by quantum mechanics ($2\sqrt{2}\approx 2.828$).  These values are on par with the highest reported violations of Bell’s inequality ever reported \cite{29}, but unlike all previously reported results  include \textit{no accidental subtraction or post-processing of any kind}.  As a result, this source provides not only the best experimental evidence to date that local realistic theories are not viable, but also provides the best test so far of the \textit{upper limits} for quantum correlations; some super-quantum theories \cite{30} actually predict that the upper limit for the Bell inequality can be \textit{greater} than $2\sqrt{2}$, a prediction constrained by the results reported here \cite{18}.

The high entanglement quality, along with the detection-loophole-free capability, offers interesting possibilities for applications, notably for ``device-independent" quantum information processing. Here the goal is to implement a certain protocol, and to guarantee its security, without relying on assumptions about the internal functioning of the devices used in the protocol. Being device-independent, this approach is more robust to device imperfections compared to standard protocols, and is in principle immune to side-channel attacks (which were shown to jeopardize the security of some experimental quantum cryptography systems).

One prominent example is device-independent randomness expansion (DIRE) \cite{31,32,33,34}. By performing local measurements on entangled particles, and observing nonlocal correlations between the outcomes of these measurements, it is possible to certify the presence of genuine randomness in the data in a device-independent way.  DIRE was recently demonstrated in a proof-of-principle experiment using entangled atoms located in two traps separated by one meter \cite{31}; however, the resulting 42 random bits required a month of data collection!  Here we show that our setup can be used to implement DIRE much more efficiently. The intrinsic randomness of the quantum statistics can be quantified as follows. The probability for any observer (hence, for any potential adversary) to guess the measurement outcome (of a given measurement setting) is bounded by the amount of violation of the CH inequality: $p_{guess}\le (1+\sqrt{2-(1+2B)^{2}})/2$ \cite{31}, neglecting finite size effects.  In turn, this leads to a bound on the min-entropy per bit of the output data, $H_{min}=-log_{2}(p_{guess})$. Finally, secure private random bits can be extracted from the data (which may in general not be uniformly random) using a randomness extractor \cite{35}. At the end of the protocol, a final random bit string of length $L\approx N*H_{min}-S$ is produced, where $N$ is the length of the raw output data, and $S$ includes the inefficiency and security overhead of the extractor.

Over the 4450 measurement blocks (each block features 25,000 events), we acquire 111,259,682 data points for 3 hours of data acquisition. The average CH violation of $B = 5.4$x$10^{-5}$ gives a min-entropy of $H_{min} = 7.2$x$10^{-5}$.  Thus, we expect \url{~}8700 bits of private randomness, of which one could securely extract at least 4350 bits \cite{18}.  The resultant rate (0.4 bits/s) improves by more than 4 orders of magnitude over the bit rate achieved in \cite{31} ($1.5$x$10^{-5}$ bits/s). This shows that efficient and practical DIRE can be implemented with photonic systems.

We have presented a new entangled photon pair creation, collection, and detection apparatus, where the high system efficiency allowed us to truly violate a CH Bell inequality with no fair-sampling assumption (but still critically relying on the no-signaling assumption that leaves the causality loophole open).  Because photonic entanglement is particularly amenable to the types of fast, random measurement and distributed detection needed to close the locality loophole, this experiment (together with efforts by other groups \cite{8,12,15}) represents the penultimate step towards a completely loophole-free test of Bell’s inequality, and advanced device-independent quantum communication applications; here we demonstrated production of provably secure randomness at unprecedented rates.  Finally, our high source quality enables the best test to date of the \textit{quantum mechanical} prediction itself.

This research was supported by the DARPA InPho program: US Army Research Office award W911NF-10-1-0395, NSF PHY 12-05870, the NIST Quantum Information Science Initiative, the Swiss NSF ($\#$PP00P2 138917), the EU chist-era project DIQIP and the Swiss NCCR-QSIT.  The data reported in this paper is available on our website at http://research.physics.illinois.edu/QI/Photonics/BellTest.

The authors of this paper also acknowledge helpful discussions with Manny Knill.

Corresponding author: Bradley Christensen (bgchris2@illinois.edu).



$\\$

\section{\label{sec:level0}Supplementary Information for: Detection-Loophole-Free Test of Quantum Nonlocality, and Applications}
\noindent{Contains: Detailed system parameters, Background counts, Estimation of error, Randomness extraction, Constraint on super-quantum models, Loopholes of previous photon experiment, and Intuition about detection-loophole and non-maximally entangled states.}

\section{\label{sec:level2}Detailed system parameters}

We use a mode-locked frequency-tripled Nd:YAG laser (120-MHz repetition rate, 5-ps pulse width, $\lambda$ = 355 nm).  A high-power PBS and HWP attenuate the laser power (from 4 W to \url{~}20 mW) to reduce the amount of pairs produced per second (multi-photon downconversion events harm the Bell violation, and the problem is greatly exacerbated by the 1-$\mu s$ coincidence window to accommodate the large TES jitter).  As the TES detector jitter is also much larger than the interpulse spacing of our pulsed laser ($1~\mu s$ compared to 8.3 ns), we use a BBO Pockels cell (Extinction Ratio (ER) $>300:1$) followed by a standard birefringent polarizer to create 2-$\mu s$-long pulse trains at a 25 kHz rate (approximately 240 laser pulses).

Our type-I phase-matched spontaneous parametric downconversion source \cite{21,22} consists of a pair of BiBO crystals, cut at $\theta = 141.8^{o}, \phi = 90^{o}$, and $\gamma = 0^{o} (90^{o})$ for the first (second) crystal.  Each crystal is only 200-$\mu$m thick to reduce the birefringent walkoff of the two downconversion beams (\url{~}$20~\mu m$).  We precompensate the temporal walkoff \cite{36} with a 550-$\mu$m piece of BBO (cut at $\theta =90^{o}$), resulting in very high entanglement quality (an interference visibility $>99.5\%$, i.e., how well the $|HH \rangle$ and $|VV \rangle$ terms interfere in an arbitrary basis).  The pump is focused to a 350-$\mu m$ radius waist on the crystal, while the downconversion is collected from a 60-$\mu m$ waist, which is located near the crystals.  To image the downconversion crystal spot onto the collection fiber, we use a 250-mm focal length plano-convex lens with an 11-mm focal length asphere on each arm of the downconversion.  All lenses are anti-reflection coated.  With this setup, we see a 90\% spatial heralding efficiency.

The basis selection is completed using a standard HWP at 710 nm, with a custom-coated Brewster’s angle PBS (ER $>$ 8700:1, T $> 99\%$) to make the projective measurement.  Finally, we use a combination spectral filter, comprised of two interference filters (see Fig. S1).  One of the filters has a lower bandedge at \url{~}700 nm (this was custom set at the company by heat treating the filter), while the other filter has an upper bandedge that can be tuned by tilting the filter.  We then tilt these filters to maximize the heralding efficiency, which will set the upper bandedge to \url{~}720 nm (the wavelength of the upper edge for the signal photon must be conjugate to the lower edge of the idler photon and vice versa, i.e. $\frac{1}{\lambda_{signal,upper}}=(\frac{1}{\lambda_{pump}}-\frac{1}{\lambda_{idler,lower}})^{-1}$).  We were thus able to achieve a spectral heralding efficiency (the heralding efficiency if there is no other loss in the system) of $95\%$.  After accounting for all losses, the total heralding efficiency is \url{~}76\% from the $|HH \rangle$ crystal, and \url{~}69\% efficiency from the $|VV \rangle$ crystal (due to birefringent transverse walkoff).  However, because the measurements are made at angles near $H$, we estimate the net heralding efficiency to be \url{~}75\%.

\begin{figure}
\includegraphics[scale=0.23]{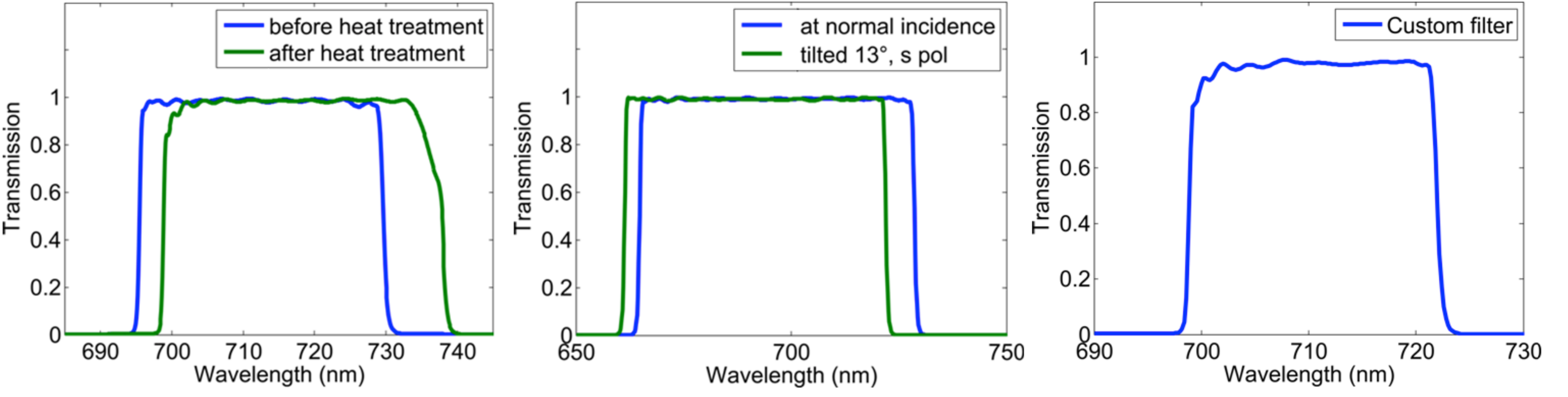}
\caption{\label{fig:first} The combined filter spectrum.  The lower bandedge (at 700 nm) is set by a fixed filter, while the upper bandedge (at 720 nm) is set by a tunable filter.  The upper bandedge is controlled by tilting the tunable filter, and adjusted to maximize the heralding efficiency.  The resultant spectral heralding is $95\%$.}
\end{figure}

\section{\label{sec:level2}Background counts}
The fluorescence rate (broadband emission believed to be mostly from the anti-reflection coating on the BiBO crystal surfaces) was measured by pumping the crystal with only an H(V)-polarized pump, and orienting the downconversion photon polarizers to H(V) as well.  By accounting for the extinction ratio of the polarizer, we can estimate the number of unpolarized broadband (i.e., not downconversion) photons produced from the crystal.  We observed rates ranging from $0.15\%$ to $0.25\%$ of the singles counts (i.e., if we had 10,000 singles counts per second, measured with V-polarizers when the pump was H-polarized, we would see between 15-25 fluorescence counts per second, after H-polarizers).  The rate would vary depending on the spot of the crystal, and also increased over time as the coating on the crystal surface started to degrade under exposure to the UV laser.  In addition, we also detected approximately 8 cps of stray photons coupled into the SMF, i.e., even with the pump laser off.

\section{\label{sec:level2}Estimation of error}
To estimate the error of the Bell parameter, we segment our data into 50 partitions by taking every 50th trial (the first partition is trials {1,51,101,…}, the second partition is trials {2,52,102,...}).  We calculate the Bell parameter $B$ for each of the partitions and look at the standard deviation of the 50 Bell parameters.  To improve the statistical certainty, we estimate $p_{1}$($a$) by including both $S$($a$) terms and dividing by $N$($a,b$) + $N$($a,b'$), and similarly for $p_{2}$($b$).  We then use error propagation to estimate the uncertainty (i.e., we divide the standard deviation by $\sqrt{49}$ since we have 50 segments).  To increase the confidence of the determined uncertainty, we also performed the previous analysis using 47,48,...,53 partitions.  The end result is $\sigma_{B} = 7.0$x$10^{-6} \pm 3.5$x$10^{-7}$, where $7.0$x$10^{-6}$ is the mean of the 7 uncertainties, and $3.5$x$10^{-7}$ is the standard deviation of the 7 uncertainties.

Finally, we repeated the previous analysis, except with sequential partitions (i.e., the first partition is the first $2\%$ of the data, the second partition is the second $2\%$ of the data).  Here we see $\sigma_{B} = 6.8$x$10^{-6} \pm 6.7$x$10^{-7}$, in close agreement with the non-sequential partitioning.

Another potentially more relevant approach to estimating the error for Bell tests, is to calculate the probability that a ``hacker" could have guessed the settings well enough to produce a violation.  For example, there is no reason to believe that the statistics of a violation follow a gaussian distribution (especially when one assumes some sort of hacker trying to fake the Bell test).  One possible method, which will give a definite upper bound on that probability, is to partition the data into distinct cycles and examine how many of those partitions produced a Bell violation.

In our case, we can make \textit{up to} 1086 distinct cycles (since each measurement setting was used at least 1086 times).  We find the optimal number of partitions to use in our case is around 650 partitions, where we see 394 Bell violations, and 256 non-violations (here, we estimate $p_{1}$($a$) by considering only the $S(a|b')$ term, and likewise $p_{2}$($b$) = $S(b|a')$).  Then the probability that a hacker could succeed in producing a violation on a given cycle by guessing is 50\% (a hacker does not need to guess every measurement setting correctly, one only needs to distinguish between the $\{a,b\}$ measurement setting and the $\{a',b'\}$ measurement setting, which can be done with a 50\% chance).  Thus, we see that the probability a hacker could have produced \textit{at least} the observed 394 violations by guessing is just $p = 3.5$x$10^{-8}$ (which is a definite upper bound on the probability a hacker guessed the measurement settings).

One thing to note about this approach is that it doesn't use all of the available information to determine the probability of a hacker guessing the measurement settings.  For example, a more accurate estimate could be found by asking what the probability is that the hacker could have produced the entire \textit{distribution} of the observed 1086 Bell parameters (instead of only asking what is the probability of producing at least the \textit{number} of violations we observe).  Of course, the hacker could produce any distribution with probability $(1/2)^{1086}$, putting a definite lower bound on the probability that a hacker forced the observed violation.

\section{\label{sec:level2}Randomness extraction}
The problem of extracting a relatively small amount of randomness from a larger string has been well studied.  The usual NIST-suggested methods, i.e., ``SHA–256", perform a complicated series of operations on the original data string, outputting a shorter string with higher entropy per bit.  The output can be considered essentially fully random if the amount of entropy input is at least twice the number of bits output from the extractor \cite{37}.  For our case, this would result in 8700/2 = 4350 final secret random bits.

The SHA randomness extractors are not known to be secure—``quantum-proof" — in the sense that they do not guarantee that the extractor output is uniform with respect to the side information (possibly quantum) held by a potential adversary. More recent results \cite{38}, based on work by Trevisan \cite{39}, show that one can achieve efficient quantum-proof extraction of randomness by incorporating an auxiliary random seed).  If the seed is itself uniformly random, i.e., unbiased, then one can extract almost all of the original min-entropy (8700 in our case) of the original data string (8x$10^{8}$ bits for us); specifically, the final secure, secret output string will have $8700 - 4log_{2}(1/\epsilon)$, where $\epsilon$ is the error between the actual output random string and a truly uniform random string.  Conservatively choosing $\epsilon = 10^{-9}$, we could extract 8580 bits.  However, the required (uniform) seed length to achieve this is $O([log_{2}(8$x$10^{8})]^{3}) \approx 26000$.  Although we can generate such a seed from the same QRNG used for selecting the measurement settings in our Bell test, the relative seed overhead cost will be better reduced by running on a larger initial data string.  Note that this analysis does not attempt to take finite statistics into account, which will be the topic of future studies.

\section{\label{sec:level2}Constraint on super-quantum models}
There are two simple super-quantum models that are easy to consider.  The first model is one in which the maximum CHSH Bell parameter is greater than $2\sqrt{2}$.  In this case, our measured CHSH Bell parameter of $2.827 \pm 0.017$ sets the upper limit of the allowable super-quantum value to 2.861 (at $2\sigma$).  The other case to consider would be an ``on/off" model, where the super-quantum model (which is assumed to saturate the Bell violation to $S=4$) only exists a percentage of the time; here, we find that such a super-quantum model can only be ``on" for 2.8\% of the time (at $2\sigma$).  While more complex models (and analysis of these models) can exist, the results from our experiment can begin to set bounds on the strength of these super-quantum models.

\section{\label{sec:level2}Loopholes of previous photon experiment}
In this section, we will discuss the two primary loopholes of the previous reported experimental results \cite{15} (outside of the causal loophole) that require fair-sampling assumptions.  While it may be possible to look at correlations in the whole data set (rather than just by looking at events) to still close the detection loophole \cite{17}, the results presented in the paper, and in the follow up paper \cite{40}, do not justify the claim that the paper is free of fair-sampling assumptions.

The first issue, called the coincidence-time loophole, was discovered by J.-A. Larsson and R. D. Gill in \cite{16}, and brought to our attention by Manny Knill \cite{17}.  Larsson and Gill discuss the loophole from the perspective of a measurement-setting-dependent detector; here we will describe it from the perspective of an adversary trying to force a Bell violation using only non-entangled photons, and without using the timing loophole.  The goal is to describe a classical source (the hidden-variable model) that nevertheless violates the CH inequality.  We exploit the fact that a coincidence count in \cite{15} is not defined by a system clock (for example, a laser pulse), and instead is defined with respect to the detection event itself.  Consider the following hidden-variable model:\\
At time $t_{1} = T$, the source emits (to Alice) a photon that is instructed with a hidden-variable: 0 if the analysis setting is $a$, 1 if $a'$.\\
At time $t_{2} = 2T$, the source emits (to Bob) a photon with instructions: 0 if $b'$, 1 if $b$.\\
At time $t_{3} = 3T$, the source emits (to Alice) a photon with instructions: 0 if $a'$, 1 if $a$.\\
At time $t_{4} = 4T$, the source emits (to Bob) a photon with instructions: 0 if $b$, 1 if $b'$.\\

Now, if we have $N$ trials where the previous model is used, and we choose a coincidence window ($W$) of radius $T<W<3T$, we will detect the following counts (see Fig. S2A):\\
$S_{1}(a) = N$\\
$S_{2}(b) = N$\\
$C_{12}(a,b) = N$\\
$C_{12}(a',b) = N$\\
$C_{12}(a,b') = N$\\
$C_{12}(a',b') = 0$.\\

{\parindent0pt Here we have assumed that every photon is detected, which could easily be realized in practice by simply using fairly bright pulses instead of single photons. The CH inequality, $p_{12}(a,b) + p_{12}(a',b) + p_{12}(a,b') - p_{12}(a',b') - p_{1}(a) - p_{2}(b) < 0$, then gives $1/N(N+N+N-0-N-N) = 1$.  Not only does this violate the inequality, it even exceeds the maximal violation ($1/2-1/\sqrt{2}\approx 0.207$) allowed by quantum mechanics! (Obviously we could also tune the LHV model to exactly mimic the quantum mechanical predictions.)}

One way that this loophole can be avoided is to no longer define coincidence counts with a window around a detection event, but instead with a window around some system clock (such as a laser pulse; see Fig. S2B).  A continuous pair-production source could then circumvent this loophole by partitioning time into segments and defining those segment as trials (in practice this is a very challenging task, however, as the amount of background counts increases with the size of the coincidence window, which will need to be very large to have the high efficiency necessary to close the detection loophole).  As long as the coincidence window is not defined around a detection event itself, then this hidden-variable model cannot violate a Bell inequality, e.g., the Bell violation again becomes a valid proof of the insufficiency of local realism to explain the correlations.

In our own experiment we cannot simply use the laser pulses directly to define the coincidence window, because they come at intervals (8.33 nanoseconds) much less than the timing jitter of the TES detectors (500 ns). Therefore, we use an additional Pockel cell and polarizer to reduce the effective pulse rate of the laser to only 25 kHz, so that the detector jitter is now much less than the inter-trial spacing ($40 \mu s$). Note that we choose to define an event as 240 sequential pulses. This number is chosen to minimize the ratio of noise counts to entangled photon counts; that is, we want to include more laser pulses to reduce the effect of detection of stray photons, but we want to limit the number of laser pulses to reduce the number of ``accidental" coincidence counts from two separate downconversion events.  We use 240 laser pulses as a compromise between the two effects.

\begin{figure}
\includegraphics[scale=0.45]{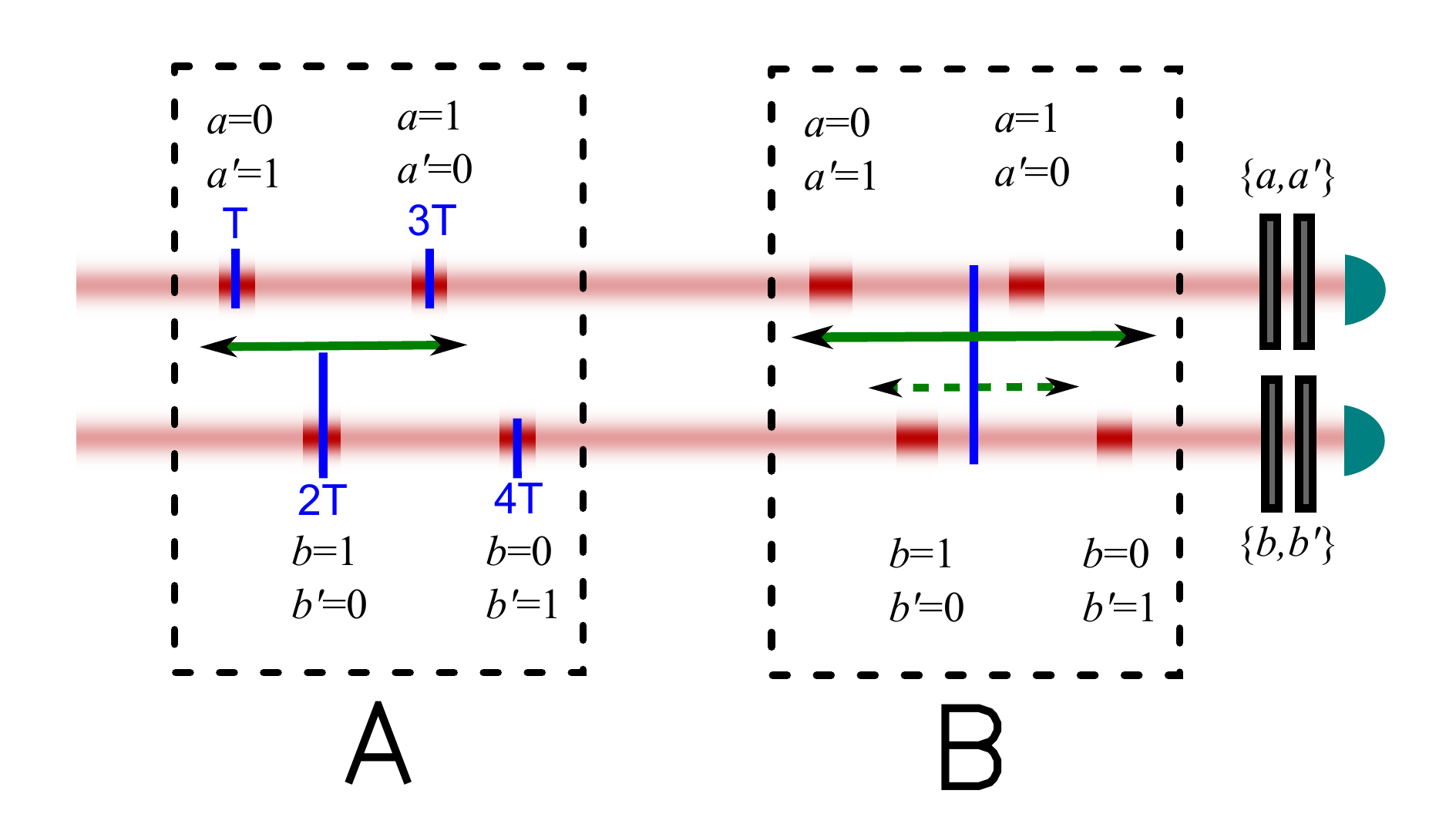}
\caption{\label{fig:second} Display of the coincidence-time loophole.  In diagram A, the coincidence window (green arrow) is defined with respect to a detection event, which allows the coincidence counts to be manipulated.  In this case, there will be no ${a',b'}$ coincidence counts, but every other coincidence count will still exist, allowing for a violation of a CH Bell inequality.  In diagram B, the coincidence window is defined with respect a system clock (blue line), instead of a detection event.  Here, the coincidence counts will not be able to be manipulated by a potential eavesdropper, and thus the source will either see coincidence counts in all channels (solid green arrow), or only ${a,b}$ coincidence counts (dashed green arrow); in either case, there will be no Bell violation.}
\end{figure}

The second weakness present in the system in \cite{15} is related to the fair-sampling assumption.  The measurements were made cyclically instead of randomly choosing the order of the analysis settings.  The reason why this is an issue can be seen in the case of a temporal-drift in laser power (either increasing or decreasing over time), or a drift in detector efficiency or latency, so that each measurement setting sees a different number of pairs produced during a count duration, i.e., the entanglement source is not fairly sampling over all of the measurement settings.  As discussed below, in principle such drift can strongly affect the Bell parameter because some of the measurements are only added negatively to the CH Bell parameter, while some measurements add positively.

Consider attempting a CH violation with the mixed state $1/2(|HH\rangle \langle HH|+|VV \rangle \langle VV|)$, which has no entanglement and therefore should not produce a Bell violation.  For this example, we will use the settings (the optimal settings if the source were maximally entangled) of $a = -b =33.75^{o}, a' = -b' = -11.25^{o}$.  Now consider making the measurements in a set order, for example: $(a,b),(a',b),(a,b'),(a',b')$.  If the laser power exponentially decays to $3\%$ of the total power over the course of the entire measurement cycle, then we would see $B' = 1.17$ (Eq. 3), corresponding to a large violation of the CH inequality.  Thus, one must make an additional assumption that that the production/detection rates are not temporally varying, and therefore the ordering of the measurements does not matter, i.e., one must assume the source is fairly sampled.  If the measurement settings are chosen based on a set of random numbers instead, as in our experiment, then these drifts affect each measurement setting equally likely over the course of many measurements.  In this case, any drift will tend to decrease the Bell violation, or at best, leave it unaffected.

As a second, more realistic example, consider attempting a CH violation with the state $|\psi_{r} \rangle = (0.3|HH \rangle + |VV \rangle)/\sqrt{1+0.3^{2}}$, in a system that only has an average detection efficiency of $76.2\%$.  Now suppose the background counts in this system are actually too high to allow for a CH violation (background count rate is $0.6\%$ of the singles rate).  For this example, we will also the settings: $a = 4.4^{o}, a' = -28^{o},b = -5.4^{o}, b' = 25.9^{o}$.  Again, consider always making the measurement in a set order, in this case: $(a,b),(a',b'),(a,b'),(a',b)$.  Here, if the laser power exponentially decays to $82\%$ of the initial power over the course of a full measurement cycle, then we would see $B = 0.005$, corresponding to the observed violation of the CH inequality, despite the noise being too high to legitimately have such a violation.

While the previous two examples show how an intensity drift could lead to a violation, the recent discussion found in \cite{40} shows that such drift, which \textit{does} exist in the system described in \cite{15}, is nevertheless not enough to cause a false violation.  However, there are still other models, such as a cyclic detector latency (which leads to varying coincidences without effecting the singles, and therefore would not be refuted by the measurements in \cite{40}) could still lead to a false violation of a Bell inequality.  In addition, the very low number of basis choice (five total cycles) puts a bound on the amount of confidence in any violation.  For example, if nature (or a hacker) were to try to guess the settings and force a violation, then they have a 50\% of succeeding for each cycle.  Thus, a hacker could have forced a violation for each cycle with a probability of $(1/2)^5=0.03$, the equivalent of a 2.2-$\sigma$ violation.

\section{\label{sec:level2}Intuition about detection-loophole and non-maximally entangled states}
In this section, we will discuss a hidden-variable model that takes advantage of the detection-loophole, as well as a discussion on how non-maximally entangled states reduce the efficiency required to close this loophole.  Of course, since the CH inequality has no fair-sampling assumption, we will base this discussion on the CHSH inequality (where we can have a fair-sampling assumption).  For each pair in a standard CHSH Bell inequality’s experimental ensemble, either measurement $a$ or $a'$ is made on particle $A$ and similarly $b$ or $b'$ on particle $B$.  The randomly chosen, relativistically simultaneous measurements result in four correlation factors $E(a_{i}b_{j})$, where $E = 1$ indicates perfect correlation (the polarizations are found to be the same), $E = -1$ indicates perfect anti-correlation (the polarization are always found to be orthogonal), and values between -1 and +1 indicate a linear mix of the two.  Bell proved that all local realistic experimental outcomes obey the following inequality:
\begin{eqnarray}
S = E(a,b) + E(a,b')+ E(a',b)- E(a',b')\leq 2
\label{eq:one}.
\end{eqnarray}
Quantum mechanics instead predicts that maximally entangled states can reach a value of $S = 2\sqrt{2}$.  In practice, each of the $E(a_{i}b_{j})$ values cannot be measured directly, but are instead inferred from a series of coincidence counts for each of the four terms above, as well as singles counts for each detector's two measurement settings.  Perhaps surprisingly, the singles counts are crucial to a loophole-free violation of local realism.  To understand why, consider inventing a local realistic explanation for the following Bell test results:

$C(a,b)= 427, C(a,b')= 427, C(a',b)= 427, C(a',b')= 73,
S(a) = 500, S(a') = 500, S(b) = 500, S(b') = 500$.

These are the average experimental results for ideal detectors, the maximally entangled state HH+VV, optimal polarizer angles of $a = -11.25^{o}, a' =33.75^{o}, b =11.25^{o},$ and $b' = -33.75^{o}$, and an ensemble of 4000 entangled pairs randomly measured in each of four configurations.  The best local realistic model for these results instructs some particle pairs to always be detected (denoted by {$a,a',b,b'$}, where the measurements listed in curly braces indicate instructions to be detected by that measurement) and some particle pairs to be detected only by a subset of all possible measurements, in the following scheme:\\
4 x 73 Pairs: {$a,a',b,b'$},\\
4 x 177 Pairs: {$a,a',b$},\\
4 x 177 Pairs: {$a,b,b'$},\\
4 x 177 Pairs: {$a,b'$},\\
4 x 177 Pairs: {$a’,b$},\\
4 x 73 Pairs: {$a'$},\\
4 x 73 Pairs: {$b'$},\\

which results in the following coincidence and singles count totals:

$C(a,b) = 427, C(a,b') = 427, C(a',b) = 427, C(a',b') = 73,
S(a) = 604, S(a') = 500, S(b) = 604, S(b') = 500$.

The local realistic model simply cannot combine multiple-basis correlations into a single set of instructions; as a result any local realistic model that explains the coincidence counts necessarily generates too many singles counts.  It is precisely this type of multiple-basis correlation that makes quantum mechanical entanglement nonclassical.

Now consider the same experiment with $82\%$-efficient detectors.  The quantum mechanical case gives:

$C(a,b) = 287, C(a,b') = 287, C(a',b) = 287, C(a',b') = 49,
S(a) = 410, S(a') = 410, S(b) = 410, S(b') = 410$.

These results are now perfectly explained by the following local realistic model:\\
4 x 49 Pairs: {$a,a',b,b'$},\\
4 x 119 Pairs: {$a,a',b$},\\
4 x 119 Pairs: {$a,b,b'$},\\
4 x 119 Pairs: {$a,b'$},\\
4 x 119 Pairs: {$a',b$},\\
4 x 123 Pairs: {$a'$},\\
4 x 123 Pairs: {$b'$},\\
4 x 4 Pairs: {$a$},\\
4 x 4 Pairs: {$b$},\\

To try to disprove this type of local realistic model, even with inefficient detectors, it is necessary to increase the coincidence-to-singles ratio, even at the expense of the absolute Bell violation (the ratio of the first three terms to the fourth in Eq. S1).  This can be accomplished with a nonmaximally entangled state, e.g., $|\psi_{r} \rangle = (r|HH\rangle + |VV\rangle)/\sqrt{1+r^{2}}$.

For $|\psi_{r} \rangle$, the optimal Bell measurement angles all collapse towards the horizontal axis: $a= -x, a'=3x, b=x$, and $b'=-3x$, where $x=11.25^{o}$ for $r=1$.  This collapse increases the coincidence-to-singles ratio, slightly increases the degree of positive correlation in the first three Bell terms, and dramatically decreases the degree of anticorrelation in the final Bell term (because the overlap between $a'$ and $b'$ grows quickly).  While the value of $S$ decreases, the experiment becomes more robust against inefficient detectors.

Note that all of the polarizer angles become close to horizontal as $x\rightarrow 0$, which reduces all coincidence and singles counts.   As a result, nonmaximally entangled states measured in this configuration become much more sensitive to noise, as shown in the Eberhard paper \cite{20}.  Physically, the problem is that the noise is unpolarized, so that the analysis settings cannot be adjusted to reduce/eliminate it, as they can for the singles counts arising from a non-maximally entangled state.


This research was supported by the DARPA InPho program: US Army Research Office award W911NF-10-1-0395, NSF PHY 12-05870, the NIST Quantum Information Science Initiative, the Swiss NSF ($\#$PP00P2 138917), the EU chist-era project DIQIP and the Swiss NCCR-QSIT.  The data reported in this paper is available on our website at http://research.physics.illinois.edu/QI/Photonics/BellTest.

The authors of this paper also acknowledge helpful discussions with Manny Knill.

Corresponding author: Bradley Christensen (bgchris2@illinois.edu).




\begin{thebibliography}{99}

  \bibitem{1}
A. Einstein, B. Podolsky, N. Rosen, Can quantum-mechanical description of physical reality be considered complete? \textit{Phys. Rev.} \textbf{47}, 777-780 (1935).

  \bibitem{2}
J. Bell, On the Einstein-Podolsky-Rosen paradox. \textit{Physics} \textbf{1}, 195 (1964).

  \bibitem{3}
S. Freedman, J. Clauser, Experimental test of local hidden-variable theories. \textit{Phys. Rev. Lett.} \textbf{28}, 938-941 (1972).

  \bibitem{4}
A. Aspect, J. Dalibard, G. Roger, Experimental test of Bell's inequalities using time-varying analyzers. \textit{Phys. Rev. Lett.} \textbf{49}, 1804-1807 (1982).

  \bibitem{5}
Z. Ou, L. Mandel, Violation of Bell's inequality and classical probability in a two-photon correlation experiment. \textit{Phys. Rev. Lett.} \textbf{61}, 50-53 (1988).

  \bibitem{6}
Y. Shih, C. Alley, New type of Einstein-Podolsky-Rosen-Bohm experiment using pairs of light quanta produced by optical parametric down conversion. \textit{Phys. Rev. Lett.} \textbf{61}, 2921-2924 (1988).

  \bibitem{7}
P. Kwiat, et al., New high-intensity source of polarization-entangled photon pairs. \textit{Phys. Rev. Lett.} \textbf{75}, 4337-4341 (1995).

  \bibitem{8}
G. Weihs, T. Jennewein, C. Simon, H. Weinfurter, A. Zeilinger, Violation of Bell's inequality under strict Einstein locality conditions. \textit{Phys. Rev. Lett.} \textbf{81}, 5039-5043 (1998).

  \bibitem{9}
W. Tittel, J. Brendel, H. Zbinden, N. Gisin, Violation of Bell inequalities by photons more than 10 km apart. \textit{Phys. Rev. Lett.} \textbf{81}, 3563-3566 (1998).

  \bibitem{10}
M. Rowe, et al., Experimental violation of a Bell's inequality with efficient detection. \textit{Nature} \textbf{409}, 791-794 (2001).

  \bibitem{11}
M. Ansmann, et al., Violation of Bell’s inequality in Josephson phase qubits. \textit{Nature} \textbf{461}, 504-506 (2009).

  \bibitem{12}
J. Hofmann, et al., Heralded entanglement between widely separated atoms. \textit{Science} \textbf{337}, 72-75 (2012).

  \bibitem{freedomofchoice}
T. Scheidl, et. al., Violation of local realsim with freedom of choice. \textit{PNAS} \textbf{107} 19708 (2010).

  \bibitem{13}
T. Marshall, E. Santos, F. Selleri, Local realism has not been refuted by atomic cascade experiments. \textit{Phys. Lett. A} \textbf{98}, 5-9
(1983).

  \bibitem{14}
I. Gerhardt, et al., Experimentally faking the violation of Bell’s inequalities. \textit{Phys. Rev. Lett.} \textbf{107}, 170404 (2011).

  \bibitem{15}
M. Giustina, et al., Bell violation using entangled photons without the fair-sampling assumption. \textit{Nature} \textbf{497}, 227-230 (2013).

  \bibitem{16}
J.-A. Larsson, R. Gill, Bell's inequality and the coincidence-time loophole. \textit{Europhys. Lett.} \textbf{67}, 707-713 (2004).

  \bibitem{17}
Private communications with Manny Knill (NIST Boulder).

  \bibitem{18}
See supporting material online.

  \bibitem{19}
J. Clauser, M. Horne, Experimental consequences of objective local theories. \textit{Phys. Rev. D} \textbf{10}, 526–535 (1974).

  \bibitem{upperlimitusingmaxentangle}
D. Mermin, Ann. N.Y. Acad. Sci. 480, 422 (1986).

  \bibitem{20}
P. Eberhard, Background level and counter efficiencies required for a loophole-free Einstein-Podolsky-Rosen experiment. \textit{Phys.
Rev. A} \textbf{47}, R747-R750 (1993).

  \bibitem{21}
P. Kwiat, E. Waks, A. White, I. Appelbaum, P. Eberhard, Ultrabright source of polarization-entangled photons. \textit{Phys. Rev. A} \textbf{60}, 773-776 (1999).

  \bibitem{22}
A. White, D. James, P. Eberhard, P. Kwiat, Nonmaximally entangled states: production, characterization, and utilization. \textit{Phys. Rev. Lett.} \textbf{83}, 3103-3107 (1999).

  \bibitem{23}
M. Wayne, E. Jeffrey, G. Akselrod, P. Kwiat, Photon arrival time quantum random number generation. \textit{J. Mod. Opt.} \textbf{56}, 516 (2009).

  \bibitem{24}
R. Bennink, Optimal collinear Gaussian beams for spontaneous parametric down-conversion. \textit{Phys. Rev. A} \textbf{81}, 053805 (2010).

  \bibitem{25}
A. Miller, S. Nam, J. Martinis, A. Sergienko, Demonstration of a low-noise near-infrared photon counter with multiphoton discrimination. \textit{Appl. Phys. Lett.}  \textbf{83}, 791 (2003).

  \bibitem{26}
A. Lita, A. Miller, S. Nam, Counting near-infrared single-photons with $95\%$ efficiency. \textit{Opt. Expr.} \textbf{16}, 3032–3040 (2008).

  \bibitem{27}
N. Gisin, H. Zbinden, Bell inequality and the locality loophole: active versus passive switches. \textit{Phys. Rev. A} \textbf{264}, 103-107 (1999).

  \bibitem{28}
J. Clauser, M. Horne, A. Shimony, R. Holt, Proposed experiment to test local hidden-variable theories. \textit{Phys. Rev. Lett.} \textbf{23}, 880-884 (1969).

  \bibitem{29}
J. Altepeter, E. Jeffrey, P. Kwiat, Phase-compensated ultra-bright source of entangled photons. \textit{Optics Express} \textbf{22}, 8951-8959 (2005).

  \bibitem{30}
S. Popescu, D. Rohrlich, Nonlocality as an axiom. \textit{Found. Phys.} \textbf{24}, 379-385 (1994).

  \bibitem{31}
S. Pironio, et al., Random numbers certified by Bell's theorem. \textit{Nature} \textbf{464}, 1021-1024 (2010).

  \bibitem{32}
R. Colbeck, A. Kent, Private randomness expansion with untrusted devices. \textit{J. Phys. A} \textbf{44}, 095305 (2011).

  \bibitem{33}
S. Pironio, S. Massar, Security of practical private randomness generation. \textit{Phys. Rev. A} \textbf{87}, 012336 (2012).

  \bibitem{34}
S. Fehr, R. Gelles, C. Schaffner, Security and composability of randomness expansion from Bell inequalities. \textit{Phys. Rev. A} \textbf{87}, 012335 (2012).

  \bibitem{35}
R. Shaltiel, Recent developments in explicit constructions of extractors. \textit{Bulletin of the European Association for Theoretical Computer Science}, \textbf{77}:67-95 (2002).

  \bibitem{36}
R. Rangarajan, M. Goggin, P. Kwiat, Optimizing type-I polarization-entangled photons. \textit{Opt. Expr.} \textbf{17} 18920 (2009).

  \bibitem{37}
E. Barker, J. Kelsey, Recommendation for the entropy sources used for random bit generation. (draft) \textit{NIST SP800-90B}, Section 6.4.2, August 2012.

  \bibitem{38}
A. De, C. Portmann, T. Vidick, and R. Renner, Trevisan’s extractor in the presence of quantum side information. \textit{SIAM Journal on Computing} \textbf{41}, 915 (2012).

  \bibitem{39}
L. Trevisan, Extractors and pseudorandom generators. \textit{Journal of the ACM} \textbf{48}, 860 (2001).

  \bibitem{40}
J. Kofler, .et al., arXiv:1307.6475v1 [quant-ph].

\end{thebibliography}
\end{document}